\documentclass[pdftex,jgrga]{agutex}
\usepackage[dvips]{graphicx}
\setkeys{Gin}{draft=false}
\authorrunninghead{DALENA ET AL.}
\titlerunninghead{oxygen and bifurcated current}
\authoraddr{S. Dalena,
Dipartimento di Fisica, Universit\`a della Calabria, Rende (CS), ITALY}

\begin{document}

\title{The role of oxygen ions in the formation of a bifurcated current sheet in the magnetotail}
\author{S. Dalena$^1$, A. Greco$^{1,2}$, G. Zimbardo$^{1,2}$, P. Veltri$^{1,2}$}
\affil{$^1$Dipartimento di Fisica, Universit\`a della Calabria, Rende (CS), ITALY}
\affil{$^2$Consorzio Nazionale Interuniversitario per le Scienze Fisiche della Materia (CNISM), Unita' di Cosenza, Ponte P. Bucci, Cubo 31 C, Rende (CS), ITALY}

\begin{abstract}
Cluster observations in the near-Earth magnetotail have shown that sometimes the current sheet is bifurcated, i.e. it is divided in two layers. The influence of magnetic turbulence on ion motion in this region is investigated by numerical simulation, taking into account the presence of both protons and oxygen ions. The magnetotail current sheet is modeled as a magnetic field reversal with a normal magnetic field component $B_n$, plus a three-dimensional spectrum of magnetic fluctuations $\delta {\bf B}$, which represents the observed magnetic turbulence. The dawn-dusk electric field E$_y$ is also included. A test particle simulation is performed using different values of $\delta {\bf B}$, E$_y$ and injecting two different species of particles, O$^+$ ions and protons.
O$^+$ ions can support the formation of a double current layer both in the absence and for large values of magnetic fluctuations ($\delta B/B_0 = 0.0$ and $\delta B/B_0 \geq 0.4$, where B$_0$ is the constant magnetic field in the magnetospheric lobes). 
\end{abstract}

\begin{article}

\section{Introduction}
The magnetotail current sheet, which separates the northern from the southern lobe, is one of the key regions of magnetospheric physics. In the simplest 1-D approximation, it may be described by the Harris solution, an equilibrium solution of the Vlasov equation, in wich the current density maximum is near $z=0$, where the magnetic field is equal to zero. Although spacecraft observations confirmed the Harris model as a zero-order description of the current sheet, later studies and observations by spacecraft revealed a more complex structure. Previously, current density profiles with a double peak were observed in the near-Earth tail by ISEE 1 and 2 \citep{SergeevEA93} and were deduced by a statistical analysis of Geotail data in the distant tail \citep{HoshinoEA96}. More recently, bifurcated current sheets have been reported with Cluster data at 20 R$_E$ downtail \citep{NakamuraEA02, RunovEA03b}. In some of the analyzed events, the formation of the current double layer was associated with magnetic reconnection, as shown by \citet{RunovEA03a} and \citet{AsanoEA04}. Observations of 29 August 2001 and 26 September 2001, analyzed respectively by \citet{RunovEA04} and \citet{SergeevEA03}, reported on a bifurcated current sheet during apparent flapping motion of the plasma sheet, that lasted for $15$ min. The authors concluded that the bifurcation was not associated with reconnection process but was the result  of an aging process of a thin current sheet, in which non-adiabatic motion of ions results in a weaker current in the central plasma sheet, as demonstrated theoretically by \citet{ZelenyiEA02}. However, the physical origin of bifurcated current sheets remain controversial.

Contemporary to the observations, many theoretical models have been proposed.
\citet{ZelenyiEA02,ZelenyiEA03} considered the structure of a thin current sheet in presence of non-adiabatic particles, wich reduce the current at the center of the current sheet and the bifurcation has scale size of an ion gyroradius (hundreds of kilometers). \citet{SitnovEA03}  obtained bifurcated current sheets in the case of ion temperature anisotropy $T_{\perp}>T_{\parallel}$ (where $T_{\perp}$($T_{\parallel}$) is the ion temperature perpendicular (parallel) to the magnetic field), together with relatively flat current density profiles. \citet{RicciEA04} developed a kinetic simulation of a Harris current sheet in which the lower-hybrid drift instability caused the current sheet to bifurcate. \citet{DelcourtEA04} examined the behavior  of charged particles with adiabaticity parameter $\kappa$=1 in a simple current sheet model and found that the non linear dynamics lead to bifurcated current sheets on the scale of ion gyroradius.

In most previous studies, the presence of magnetic turbulence in the near and distant magnetotail was not considered. Although the magnetic turbulence is stronger during active geomagnetic periods, it is non-negligible even during the quiet times. Usually, the turbulence is stronger in the center of plasma sheet \citep{BauerEA95, HoshinoEA96}, while a more ordered magnetic configuration is found in the vicinity of the magnetospheric lobes. \citet{VeltriEA98} investigated the effect of turbulent magnetic fields on the proton dynamics for the distant magnetotail, where the normal component of the unperturbed magnetic field, $B_{0z}=B_n$, is statistically negligible (in the GSM coordinate system). By a test particle simulation, they found that the magnetic fluctuations play the role of effective scattering mechanism and the current sheet splits in two layers for  perturbation levels $\delta B/B_0 > 0.2$.
Subsequently, \citet{GrecoEA02} and \citet{ZimbardoEA03}, analyzing the near-Earth magnetotail, where the average $B_n$ is not zero and northward oriented, found that the normal magnetic field component $B_n$ and magnetic fluctuations $\delta B/B_0$ have opposite effects on the current structure and on the proton heating. Indeed, a large value of $B_n$ inhibits the $y$-motion in the quasi-neutral sheet, while magnetic fluctuations favour the $y$ motion at some distance from the center; the strong magnetic turbulence in the center of the current sheet slows down the proton motion and causes the double humped profiles of the current and ion velocity. More recently, \citet{GrecoEA07} have studied the equilibrium of the Earth magnetotail, injecting protons and electrons. A stationary three-dimensional kinetic-fluid code with protons represented by particles and electrons by a mass less fluid was developed. For a specific set of magnetotail parameters, the electron finite Larmor radius and the electron drift term are responsible for the formation of a double peak in the total current density, even in those case where the proton current density does not display any bifurcated structure and without turbulence.

A number of experimental studies show that in many cases the electrons are the main current carriers \citep{KistlerEA05, RunovEA06, IsraelevichEA08}. However, some Geotail observations show that in some periods the ion current prevails, while, in other periods, the electron current prevails \citep{ AsanoEA03}. Keeping in mind that under different magnetotail conditions the main current carriers can be either ions or electrons, here we try to assess the relative contribution to the cross tail current of different ion species, like protons and oxygen ions.
The principal motivation to study O$^+$ ions dynamics is the observation of rather energetic population of such ions, that shows a drift motion from dawn to dusk in the magnetic tail. The same observations \citep{ WilberEA04, KistlerEA05, CaiEA08} have shown that the $O^+$ behavior is very different from that of the protons H$^+$, analyzed till now, because of the different Larmor radius ($\rho_{O^+}=4\rho_{H^+}$, with the same initial energy), so that nongyrotropic effects are much more important for heavy ions. Out of the neutral sheet O$^+$ ions and protons perform circular orbits around the magnetic field lines. Nevertheless, since their Larmor radius is greater than the half thickness of the current sheet, the O$^+$ ions can reenter in the neutral sheet, performing meandering orbits. An interesting situation arises  where there is a thin current sheet and the plasma sheet is dominated by ionospheric oxygen, for example during magnetospheric active periods: these are the most likely cases to find the plasma sheet dynamics dominated by non-adiabatic ions, that may also carry a large fraction of the current. In this paper we investigate the dynamics of ionospheric O$^+$ ions in the near-Earth magnetotail in the presence and in the absence of magnetic turbulence.  We find that O$^+$ ions can support the formation of double peak even in the absence of magnetic fluctuations, as well as for relatively large values.

\section{Overview of oxygen observations}

Measurements have confirmed the presence of out-flowing ions O$^+$, H$^+$, He$^+$ and other ions in every magnetospheric region; these ions have low energy in the ionosphere and high energy in the magnetosphere, thanks to several acceleration mechanisms. It is known that the outflow of heavy ions from the ionosphere is strongly dependent on geomagnetic and substorms activity \citep{YauEA85, WilsonEA04}. In particular O$^+$ ions are not just significant in the inner magnetosphere, but actually the dominant ion species during magnetic storms \citep{Daglis91, KistlerEA05}: they are observed to stream from dawn to dusk across the tail, carrying about $5-10 \%$ of the cross-tail current. Indeed, during storm times, O$^+$ can dominate both the pressure and the density in the plasma sheet, that is already oxygen-rich because of the contribution from ion-outflow.
 
Inspection of Cluster data shows that often the formation of the double current sheet is in association to the increase of the oxygen amount in the near magnetotail, for example during the events on 1 October 2001 \citep{RunovEA03a, WilberEA04, WygantEA05, KistlerEA05}, 17 August 2001 \citep{SauvaudEA04, EcherEA08} and 15 September 2001 \citep{CaiEA08}. Generally, these layers are very thin, with typical thickness ranging between 2000 and 5000 km.

On 1 October 2001 at (0947-0951)UT, during a storm-time substorm, Cluster was located at [$-16.5; 8.0; 0.5$] R$_E$ (GSM). During this interval the Cluster barycenter crossed the neutral sheet $11$ times; at 09:48UT and at 09:50UT the current density showed a bifurcated structure \citep{RunovEA03a}. At the same time a declining trend in proton density from $0.3$ to $0.02$ cm$^{-3}$ is observed, that results in O$^+$ dominance ($n_{O^+}\simeq 0.07 cm^{-3}$) \citep{WygantEA05, WilberEA04}. This situation is rather common: \citet{KorthEA02} have shown that O$^+$/H$^+$ energy density ratio of storm-time substorms exceeds $100\%$; for non-storm substorms the ratio is less and amounts to $(15-65)\%$. Magnetic field value in the magnetospheric lobes and normal component of magnetic field are approximately 20 nT and 5 nT. We have also estimated a magnetic fluctuations level $\delta {\bf B}$ of around 10 nT. The current sheet half-thickness measures approximately 2500 km \citep{KistlerEA05}.

The event of 17 August 2001 is another example  in which the Cluster spacecraft were in the plasma sheet during a geomagnetic storm and observed the O$^+$ outflow from the ionosphere into the tail \citep{SauvaudEA04, EcherEA08}. In addition, during the storm main-phase, a substorm occurred. Oxygen beams were present in both the lobe and plasma sheet prior to this substorm, so this is an event where the plasma sheet is oxygen rich prior the substorm onset. After the substorm onset at 16:00 UT, the H$^+$ density and pressure decreased significantly, while O$^+$ pressure remained high. From 16:36 to 16:46 UT, O$^+$ was the dominant ion in the plasma sheet and showed a duskward motion.

During the Cluster spacecraft crossing  on 15 September 2001, both an embedded proton (that manifests a pressure anisotropy mainly with $p_{\parallel }> p_{\perp}$) and a bifurcated oxygen ion thin current sheet (that exhibits a pressure anisotropy mainly with  $p_{\parallel }< p_{\perp}$ and nongyrotropy) were observed \citep{CaiEA08}. Indeed, in the time interval (04:57:45-05:00:25)UT, a localized self-consistent current sheet equilibrium of oxygen ions was observed by C1 and C4, althought in this very thin sheet the current contribution from oxygen ions is minor. Magnetic field value in the magnetospheric lobes and normal component of magnetic field are approximately 25 nT and 3 nT. The estimation of the oxygen thin current sheet thickness is about 2500 km.

\section{Test particle simulation}
The magnetic field model used in the simulation is described in full detail in \citet{VeltriEA98,GrecoEA02}. We recall here some of the main features. The considered magnetic field model consists of three terms, ${\bf B}= {\bf B}_{0x}(z) + {\bf B}_n +\delta{\bf B}({\bf r})$.
${\bf B}_{0x}(z)$ is an unperturbed, sign reversing component, directed along the Earth-Sun axis. We use the expression of a modified Harris magnetic field reversal:
$$
B_{0x}=B_0\frac{\tanh{\left(\frac{z}{\lambda}\right)}-\left(\frac{z}{\lambda}\right)\cosh^{-2}{\left(\frac{L}{2\lambda}\right)}}{\tanh{\left(\frac{L}{2\lambda}\right)}-\left(\frac{L}{2\lambda}\right)\cosh^{-2}{\left(\frac{L}{2\lambda}\right)}}
$$
where B$_0$ is the constant magnetic field in the magnetospheric lobes, $\lambda=0.25L$ is the current sheet half thickness and $L$ is the thickness of the considered magnetic field configuration (i.e. of the simulation box). ${\bf B}_n=B_n{\bf e}_z$ is the unperturbed, constant, normal component. $\delta{\bf B}({\bf r})$ corresponds to magnetic fluctuations, here represented as the sum of static magnetic perturbations:
$$
\delta{\bf B}({\bf r})=\sum_{{k},\sigma}{\delta{\bf B}({\bf k}){\bf e}_{\sigma}({\bf k})\exp{[i\left({\bf k}\cdot{\bf r}+\phi_{\bf k}^{\sigma}\right)]}},
$$
where ${\bf e}_{\sigma}({\bf k})$ are the polarization unit vectors, $\phi_{\bf k}^{\sigma}$ are random phases and
$$           
\delta B_{\sigma}({\bf k})=\frac{C}{\left(k_x^2l_x^2+k_y^2l_y^2+k_z^2l_z^2+1\right)^{\alpha/4+1/2}},
$$
where C is a normalization constant and $l_x$, $l_y$ and $l_z$ are the turbulence correlation lengths in the $x$, $y$ and $z$ directions, respectively.
The correlation lengths are fixed in terms of the thickness of simulation box as $l_x=l_y=0.25L$ and $l_z=0.05L$, in order to mimic the geometry of the magnetotail (i.e. $l_z\gg l_x,l_y$). The spectral index $\alpha$ for the near-Earth magnetotail is chosen as $\alpha=2.3$ \citep[e.g.][]{ZimbardoEA09}.

 In addition to the above magnetic field configuration, we consider a constant cross tail electric field in the dawn to dusk direction, ${\bf E}_y=E_y{\bf e}_y$, that cannot be removed by transforming to the de Hoffman-Teller frame for the presence of three-dimensional magnetic fluctuations. Since we are considering static magnetic perturbations, the fluctuating electric field is not included in the present runs. We may show that the fluctuating eletric field has a minor inpact on particle dynamics, because the ratio $V_A/v_{th} \ll 1$ \citep{VeltriEA98, GrecoEA02}. A study of time dependent fluctuations and the associated energization has been carried out by \cite{GrecoEA09, PerriEA09}. Instead in this study we are interested in studying the particles dynamic and its role in the formation of the double current layer, rather than  the acceleration mechanism to which these particles could be subject. In Table \ref{tab1}, normalizations and typical values of the physical quantities, inferred from observations \citep{RunovEA06, WygantEA05, KistlerEA05}, are shown.

\begin {table}
\center
\begin{tabular}{lcr}
\hline
Physical quantity       &  Normalization            &  Typic value\\
\hline
Length                  &    $L$                    &  $4000$ km\\
Time                    &   $\omega_{0i}^{-1}$      &  2 s\\
Electric field          &    $E_0$                  &  $0.2$ mV/m\\
Magnetic field          &    B$_0$                  &  20 nT\\
Current density         &    $B_0/\mu_0 L$          &  $4$ nA/m$^2$\\
Velocity                &    $V_E=cE_0/B_0$         &  10 km/s\\
Temperature             &    $m_pV_E^2/k_B$         &  $1.2\times10^4$ K\\
Ion density            &    $n^*=cB_0/4\pi eLV_E$  &  $0.5$ cm$^{-3}$\\
\hline
\end{tabular}
\caption{Normalization quantities.}
\label{tab1}
\end{table}

We also defined $\delta B=\sqrt{\langle \delta {\bf B}\cdot \delta {\bf B} \rangle}$, with the average made over the simulation box, and $\delta b=\delta B/B_0$ and $b_n=B_n/B_0$. We consider that the source of the particles that are entering the current sheet from the lobes, at $z=\pm 0.5L$, is located somewhere in the magnetospheric mantle. The relatively cold ion distribution in the mantle magnetic field can be described as a shifted Maxwellian (e.g., \citep{Ashour-AbdallaEA94}):
$$
f(v_{\parallel},v_{\perp})=\left( \sqrt{2\pi} v_{th}^3\right)^{-1}\exp{\left(-\frac{(v_{\parallel}-u)^2+v_{\perp}^2}{2v_{th}^2}\right)},
$$
where $u$ is the streaming velocity along the unperturbed magnetic field and $v_{th}$ is the thermal velocity.

\begin{figure*}
\begin{center}
\noindent\includegraphics[width=35pc]{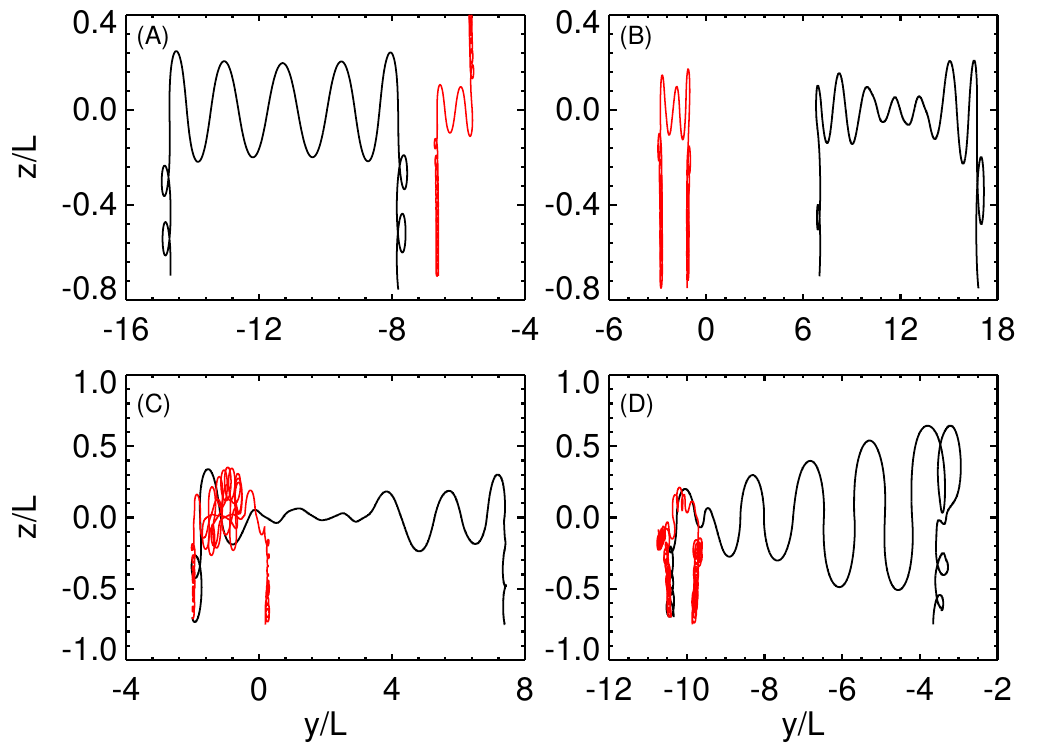}
\caption{Projection on the plane of four sample trajectories of O$^+$ ions (black line) and protons (red line) for $b_n=0.1$, $E_y=1E_0$ and $\delta b=0$ (panel (A)), $\delta b=0.1$ (panel (B)), $\delta b=0.2$ (panel (C)), $\delta b=0.6$ (panel (D)).}
\label{traj_H_O}
\end{center}
\end{figure*}

In the present simulation we assume $b_n = 0.1$ for the normal magnetic field. Typically, $500000$ particles are injected for each run with temperature $T_i=0.5$ keV, in agreement with the values observed in the magnetospheric lobes. The corresponding ion injection velocities are respectively, $v_{th}$(O$^+$)$=56$ km/s and $v_{th}$(H$^+$)$=224$ km/s \citep{VaisbergEA96}. Besides, we assume $u=v_{th}$ for both species.
In Figure {\ref{traj_H_O}}, we show the projection on the $yz$ plane of four sample trajectories of H$^+$ and O$^+$ ions, for different values of magnetic fluctuations $\delta b$.
Once entered in the simulation box, in total absence of magnetic fluctuations (panel (A)), both types of particles move toward the central region $z=0$, following the magnetic field lines while being subjected to the $({\bf E} \times {\bf B})$ drift. Inside of the quasi-neutral sheet, particles begin to perform the typical meandering orbits, under the action of the unperturbed magnetic field, while being accelerated by the electric field E$_y$. It's clear that O$^+$ has larger Larmor radius than protons of the same energy and they can carry a large fraction of the current away from the center $z=0$. The nonzero $B_n$ plays the role of a guiding channel for the particles and causes them to exit the quasineutral sheet in the $z$ direction \citep{Speiser65}. Typically the particles, after traversing variable distances in $y$, stop meandering and travel towards the lobes with larger Larmor radius, since they gained energy being accelerated by E$_y$ (this effect is more apparent in the panels B-D).

\begin{figure}
\begin{center}
\includegraphics[width=20pc]{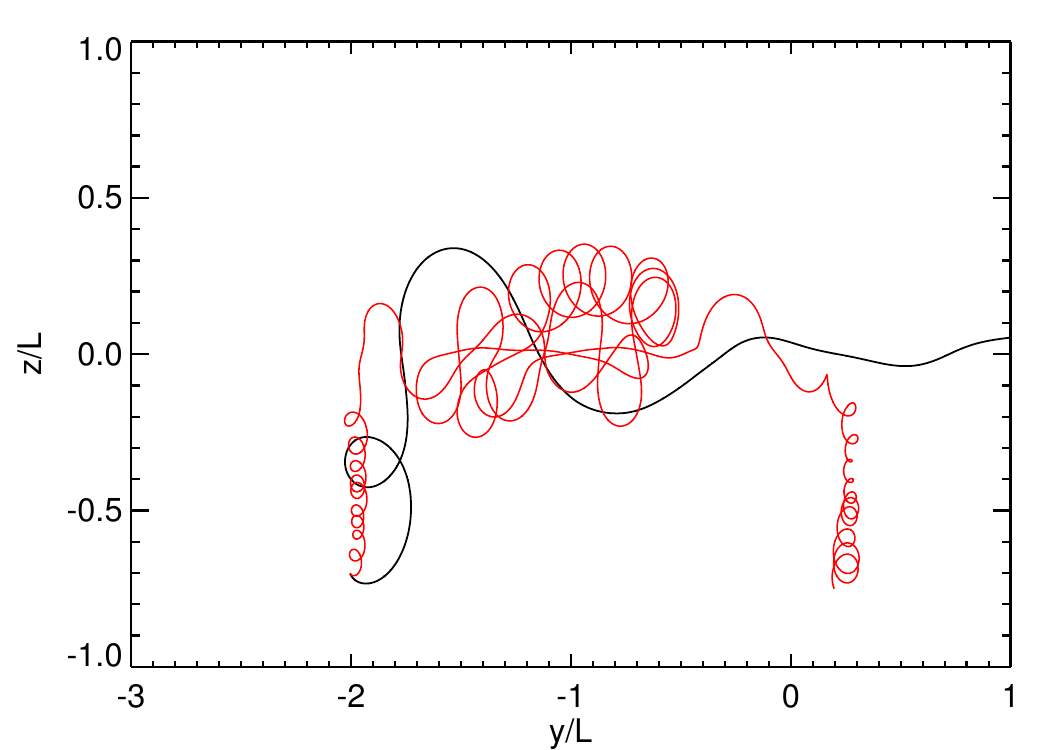}
\caption{Blow up of O$^+$ ions (black line) and protons (red line) trajectories for $b_n=0.1$, $E_y=1E_0$ and $\delta b=0.2$.}
\label{traj_p}
\end{center}
\end{figure}

Increasing the values of magnetic fluctuations $\delta b$ (panels (B), (C) and (D)), the performed orbits become are perturbed and the trajectories become more intricate, especially for H$^+$ ions (see the blow up in Figure \ref{traj_p}). In the regions of meandering motion, different deflections are observed, which slow down the motion along $y$ and make the trajectory more tangled, although this effect is more evident for protons. Starting with the same initial conditions, the O$^+$ ions interact less with the magnetic turbulence than the protons H$^+$, because of their greater Larmor radius: for $T=0.5$ keV and for $B=0.1 B_0$ (e.g. in the neutral plane), $\rho$(H$^+$) $\simeq 1130$ km and $\rho$(O$^+$) $\simeq 4600$ km.
The turbulence characteristic lengths are $\lambda_{max} = L/k_{min} = 4000$ km and $\lambda_{min} = L/k_{max} = 333$ km, where $k_{max}=12$ and $k_{min}=1$ are the maximum and minimum wave number in each direction, respectively. The O$^+$ ions are not influenced by the magnetic turbulence so much as protons; besides oxygen ions are able to transport current out of the neutral sheet, because of the amplitude of performed orbits. A similar difference in the proton and oxygen interaction with magnetic turbulence was reported by \cite{TaktakishviliEA07}, with regard to plasma transport across the magnetopause current sheet. 

In order to gain understanding in the ion dynamics, we inject many particles in the simulation box and numerically integrate their equation of motion. Then we compute the distribution function moments as density $n$, current density ${\bf j}$ and temperature $T$ on a three-dimensional grid with $1$ grid point in $x$ (many grid points in $x$ could be used; however for the present runs we set this to 1, since $x$ is a statistically ignorable coordinate in our model), $601$ in $y$ and $40$ in $z$. Then the distribution function moments have been averaged over $y$, the dependence on which is usually weak, to show the characteristic dependence on the $z$ coordinate. In a test particle simulation the normalization for the number density is, to good extent, arbitrary. Here it is based on the consideration that the ion current $I_y$ has to be strong enough to reproduce the unperturbed magnetic field $B_{0x}(z)$ introduced in our model. Indeed, Ampere's law leads to $2B_0L_x=4\pi I_y/c$, with $I_y = \int j_y dxdz$ the total current across a section of the current sheet of length $L_x$. On the other hand, velocity is expressed in units of $V_E$, so that the normalization for density is obtained an $n^*= I_y/[e \int V_ydxdz]$, with $I_y$ constant for all the runs (this implies that, when the average value of $V_y$ is large, the density is low).

\begin{figure}
\begin{center}
\includegraphics[width=21pc]{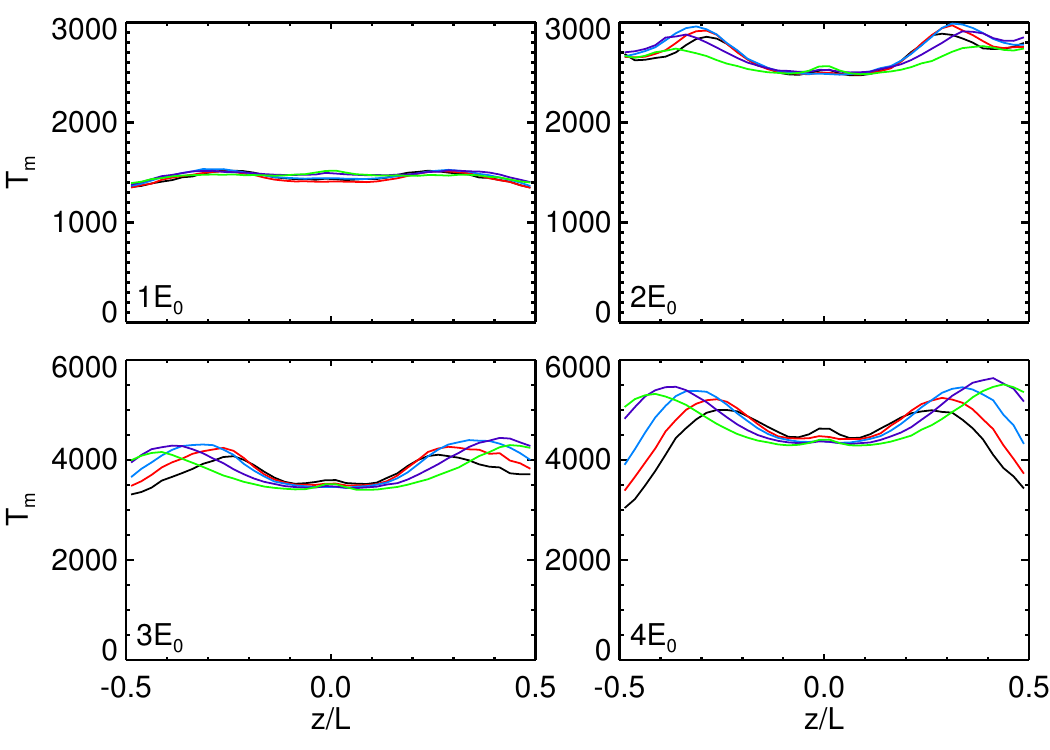}
\caption{Vertical profiles for O$^+$ ion temperature for $E_y = (1,2,3,4)E_0$ and $b_n = 0.1$. Here $\delta b = 0.0$ (black line), $\delta b = 0.1$ (red line) e $\delta b=0.2$ (blue line), $\delta b = 0.4$ (purple line), $\delta b = 0.6$ (green line). Temperatures are in units of $m_pV_E^2/k_B$.}
\label{T_O_+}
\end{center}
\end{figure}

\begin{figure}
\begin{center}
\includegraphics[width=21pc]{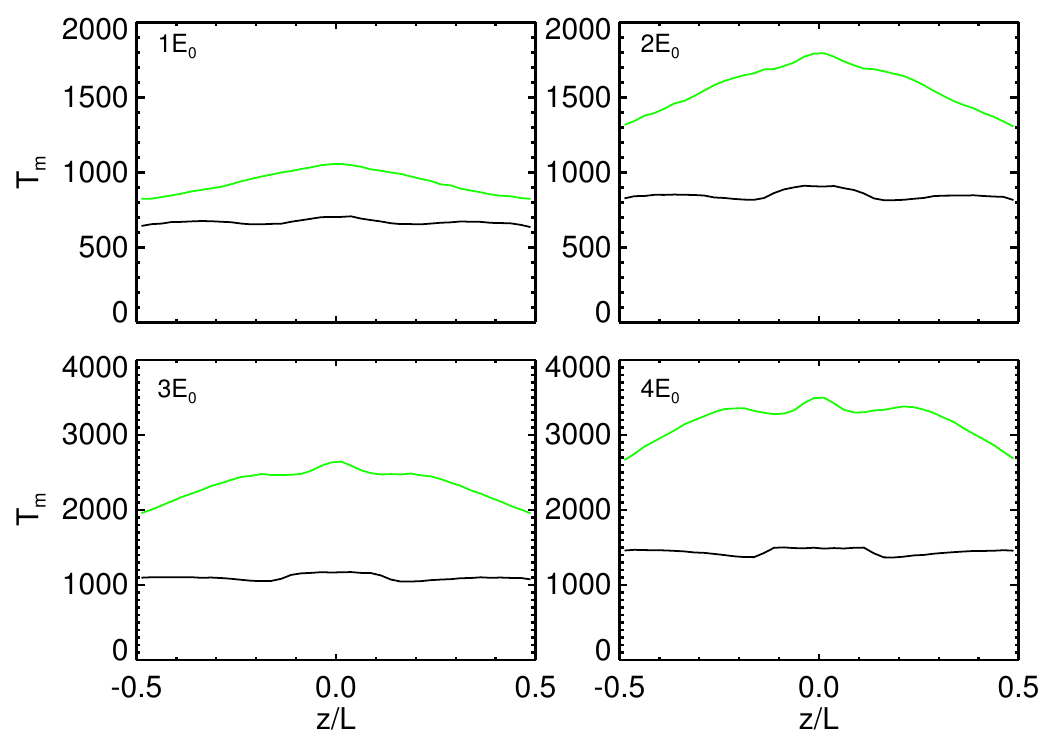}
\caption{Vertical profiles for H$^+$ ion temperature for $E_y = (1,2,3,4)E_0$ and $b_n = 0.1$. Here $\delta b = 0.0$ (black line) and $\delta b = 0.6$ (green line). Temperatures are in units of $m_pV_E^2/k_B$.}
\label{T_H_+}
\end{center}
\end{figure}

\begin{figure*}
\begin{center}
\includegraphics[width=39pc]{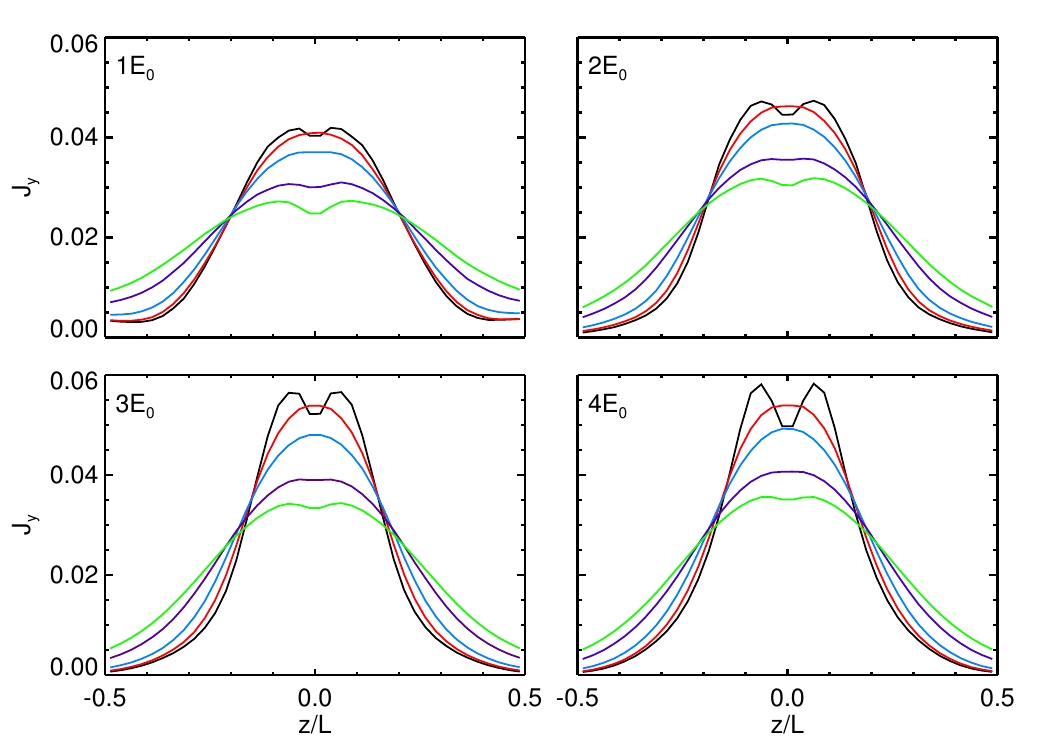}
\caption{Vertical profiles for O$^+$ ions current density for $E_y = (1,2,3,4)E_0$ and $b_n = 0.1$. Here $\delta b = 0.0$ (black line), $\delta b = 0.1$ (red line) e $\delta b=0.2$ (blue line), $\delta b = 0.4$ (purple line), $\delta b = 0.6$ (green line).}
\label{vertical_prof_bn01}
\end{center}
\end{figure*}

\begin{figure}
\begin{center}
\includegraphics[width=21pc]{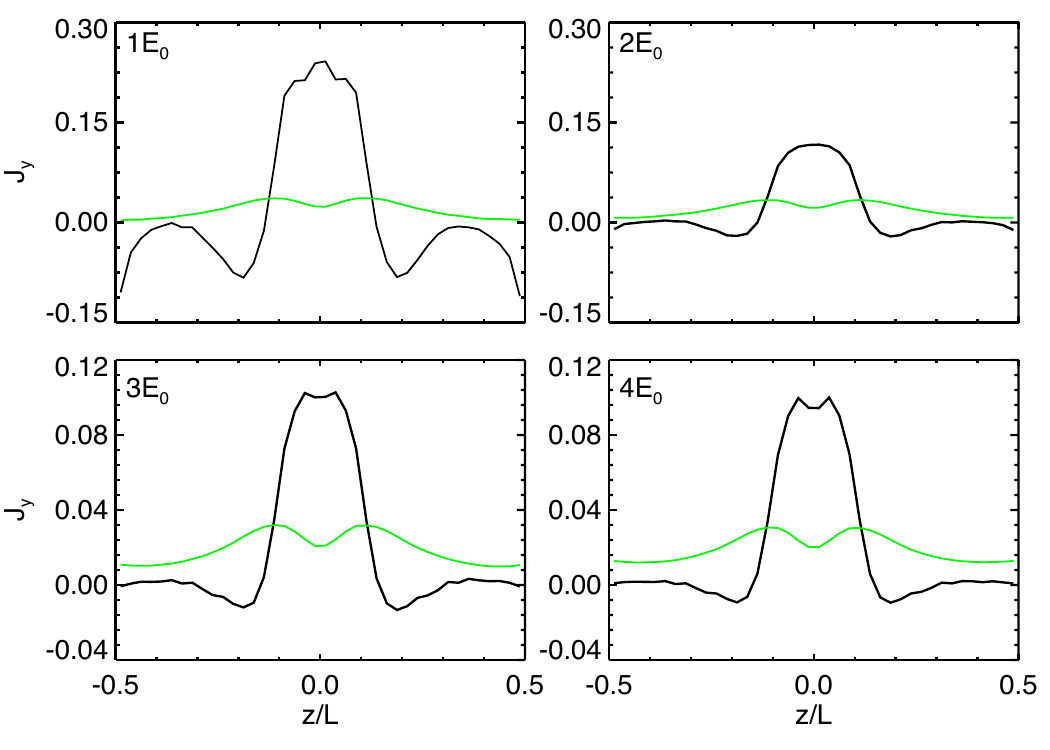}
\caption{Vertical profiles for H$^+$ ions current density for $E_y = (1,2,3,4)E_0$ and $b_n = 0.1$. Here $\delta b = 0.0$ (black line) and $\delta b = 0.6$ (green line).}
\label{j_H_+}
\end{center}
\end{figure}

In figures \ref{T_O_+} and \ref{vertical_prof_bn01}, we report the vertical profiles of oxygen temperature T and current density $J_y$, for different values of electric field $E_y = (1, 2, 3, 4)E_0$ \citep{CattellEA82}, magnetic fluctuations $\delta b = 0.0$ (black line), $0.1$ (red line), $0.2$ (blue line), $0.4$ (purple line), $0.6$ (green line) and for $b_n = 0.1$, wich corresponds to a variety of observations in the near-Earth magnetotail. In figures \ref{T_H_+} and \ref{j_H_+}, we show the vertical profiles of proton temperature T and current density $J_y$ for $E_y = (1, 2, 3, 4)E_0$,  $b_n = 0.1$ and only for two values of the magnetic fluctuations $\delta b = 0.0$ (black line), $0.6$ (green line)(results for more values of $\delta b$ have been given by \cite{GrecoEA02}). Given the abundance of previous work on protons, we restricted to show for this specie only run in which the oxygen shows a bifurcated current sheet, to be able to get the total current. The oxygen temperature in figure \ref{T_O_+} increases with the electric field, but there is no clear influence of magnetic fluctuations on heating. Indeed, in each panel, different colour lines are at same level. A different situation arises with protons. In this case, turbulence succeeds in scrambling the ordered velocity gained because of $E_y$ into all directions, leading to an effective thermalization of the potential drop \citep{GrecoEA02}. The temperature profiles are larger in the center of the simulation box, where the energization from the electric field is favored and the magnetic fluctuations are stronger. Finally, oxygen and proton temperatures are of the same order ($\sim 3 - 5$ keV) in the case of $E_y=4E_0$ and $\delta b=0.6$ (green line), in agreement with observations.

From figure \ref{vertical_prof_bn01}, it is clear that oxygen current profile is peaked and thin for zero to low fluctuation levels, and becomes progressively broader as the fluctuation level is increased, while, at the same time, the maximum current value decreases. An obvious effect of magnetic fluctuations is, indeed, to reduce the bulk velocity (and current density), by inducing random motions, which spread all around single particle velocities. Besides, oxygen current density assumes values which grow with the electric field. In the absence of magnetic fluctuations (black line), the oxygen ions are able to produce the double current layer, more and more evident when the electric field value grows. We argue that this is due to the velocity increase, caused by the electric field acceleration, so that the oxygen gyroradius and the effect of meandering orbits increases. The double peak disappears for $\delta b = 0.05$ (not shown), $0.1$ (red line), $0.2$ (blue line), and appears again for levels $\delta b \geq 0.4$ (purple and green line). The proton behaviour is similar in some aspects and opposite in other ones. We can see from figure \ref{j_H_+} that the proton current density is larger than that of the oxygen one and, in contrast with the preceding case, it decreases with increasing electric field. This is due to the presence of the negative wings, that are more evident for small values of the electric field: indeed, since the integral $I_y$ of the ion current must remain constant,  if the contribution of negative wings is big, the positive current must increase. A small hint of a double peak, in absence of magnetic fluctuations (black line), is present only for high values of electric field; instead, in presence of high magnetic fluctuations ($\delta b=0.6$ (green line)), protons support the formation of double layer for all $E_y$ values, the bifurcation being more evident with increasing of the electric field. Moreover, negative values in the current density profiles are observed: they are the so called diamagnetic currents, due to the magnetization current $-c\nabla \times (P_{\perp}\frac{\bf B}{|\bf B|})$, proportional to $v_{\perp}^2$ \citep{ZelenyiEA00}. If the anisotropy is weak ($v_{\parallel} \simeq v_{\perp}$), particles will give a substantial contribution to the formation of the diamagnetic wings; conversely, if there's a strong anisotropy ($v_{\parallel} \gg v_{\perp}$), this contribution is small \citep{GrecoEA02, ZimbardoEA04}. 

\begin{figure*}
\begin{center}
\includegraphics[width=34pc]{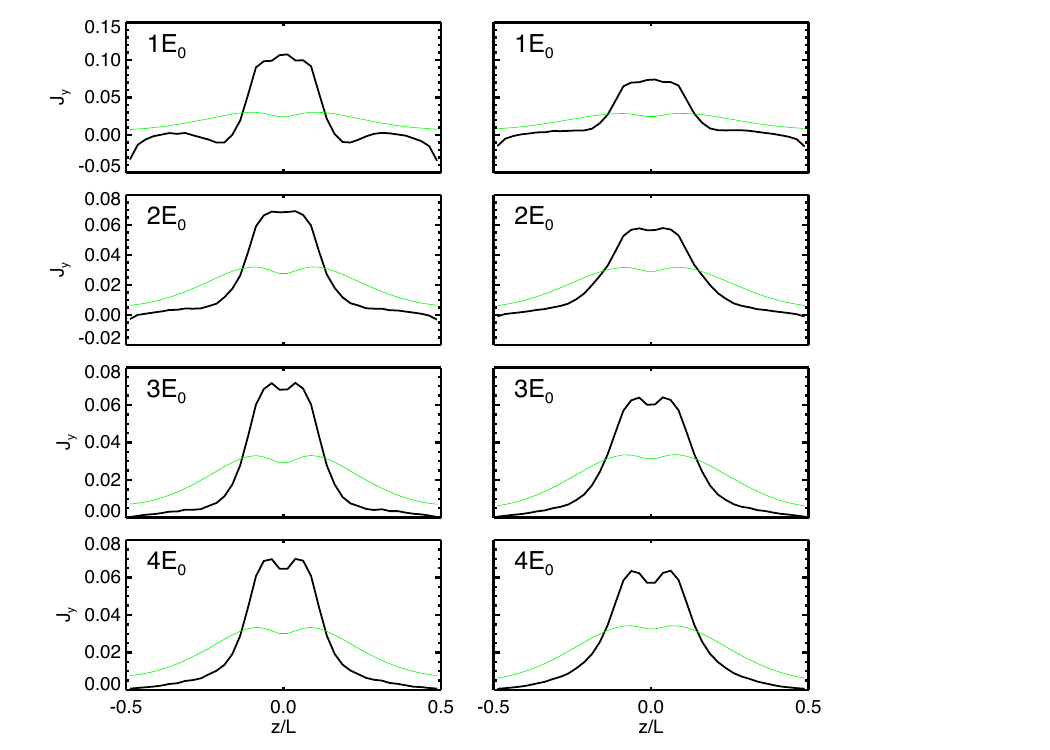}
\caption{Total current density profiles for $E_y=(1,2,3,4)E_0$, $b_n=0.1$.Here $\delta b = 0.0$ (black line) and $\delta b = 0.6$ (green line). The right column is for $k=5$, the left one is for $k=2$ (see text).}
\label{Jtot}
\end{center}
\end{figure*}

 In order to assess the oxygen contribution in the formation of a bifurcated current sheet in the total current density, we define the density ratio $k=n(O^+)/n(H^+)$. Then, we use two different $k$ values, wich are close to the real values observed in the near-Earth magnetotail, when the oxygen is the most abundant species, to find the total current density. If $k=2$, $J_{tot}=(2/3)J_{O^+} + (1/3)J_p$; if $k=5$, $J_{tot}=(5/6)J_{O^+} + (1/6)J_p$. As we can see in figure \ref{Jtot} , we found that the double peak is present in the total current density profiles, also when it is not present in the proton current density, that is for $E_y=2 E_0$ and $\delta b=0$ (black line).

\section{Discussion and Conclusions}
Motivated by the many observations of ionospheric oxygen O$^+$ in the Earth's magnetotail, in this paper we have considered the dynamics of protons and oxygen ions in the magnetotail current sheet, both in the presence and in the absence of magnetic turbulence. The ion motion has been studied by a test particle simulation, where particles are injected in a magnetic quasi-neutral sheet with superimposed magnetic fluctuations. Several runs have been made in order to understand the differences between the proton and the oxygen behavior, with emphasis of the possibility to obtain a bifurcated current sheet.
 
\begin{figure}
\begin{center}
\includegraphics[width=18pc]{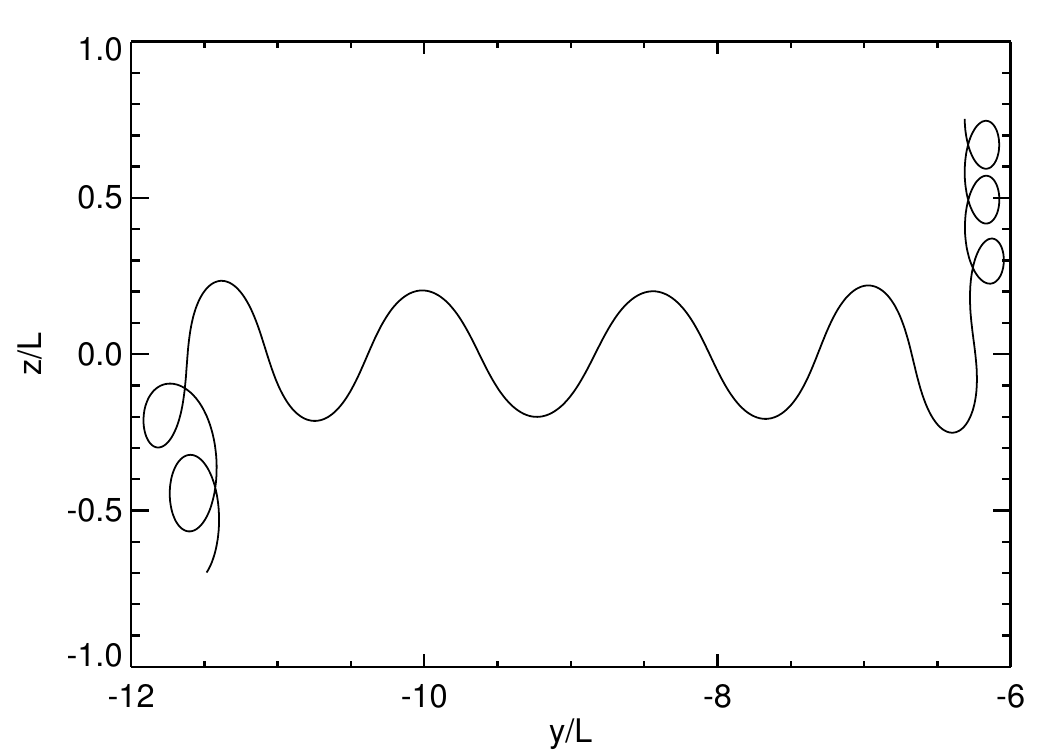}
\caption{Projection on the plane of a sample trajectory of O$^+$ ions for $b_n=0.1$, $E_y=1E_0$ and $\delta b=0$.}
\label{traj_O_deltab}
\end{center}
\end{figure}

 When protons are injected in the simulation box, a clear bifurcation of the current profile is obtained for turbulence levels $\delta B/B_0 \geq 0.3$ (see also \cite{GrecoEA02}). Conversely, because of the non adiabatic meandering motion, oxygen ions are able to support a bifurcated current sheet also in the absence of magnetic fluctuations. One of the most intriguing results concerning the oxygen behaviour is the appearance of a double hump in the current profile both for the case of no magnetic turbulence and for relatively high levels of magnetic fluctuations. The explanation for this behaviour can be found in the interaction between oxygen ions and the current sheet. Looking at figure \ref{traj_O_deltab}, which shows a typical oxygen trajectory for $\delta b=0$ and E$_y=1E_0$, we can notice that orbit is very smooth and that oxygen ion probes regions of the current sheet far from the neutral plane (because of its large Larmor radius) during its meandering motion along $y$. Therefore, this effect is essentially related to the non adiabatic motion of oxygen ions. If a great number of this kind of trajectories are statistically added, there will be a concentration of particles with high $v_y$ at those distances, obtaining a current density $J_y$ which displays two peaks away from the neutral sheet \citep{Delcourt98}. The distance between the two peaks should be of the same order of the typical excursion in $z$, that is $d \sim \sqrt {\rho_0 \lambda}$, where $\rho_0$ is the oxygen Larmor radius in lobe magnetic field $B_0$ \citep{GrecoEA02}.
From figure \ref{vertical_prof_bn01}, the separation between the two bumps for the case $\delta b=0$ (black lines) is $\sim 0.2 L=800$ km. If we consider that a typical oxygen Larmor radius, in a magnetic field of $20$ nT and with a temperature of $0.5$ keV, is of the order of $800$ km, we obtain for $d$ a value of the order of $900$ km, which is comparable with the distance between the two humps computed in the vertical profile of the current density. The influence of magnetic turbulence on $J_y$ is twofold: on the one hand, relatively low levels of magnetic perturbations ($\delta b< 0.4$) cause the two peaks to decrease and smear because fluctuations scatter the oxygen ions around. On the other hand, when the level of magnetic perturbations is high enough, the presence of fluctuations bends the field lines also in $y$ direction, and allows cross field motion away from the central plane. 

 Another important difference between the proton and the oxygen behavior is found in the temperature profiles. While the proton temperature grows with the magnetic turbulence level, the oxygen temperature does not. More precisely, the oxygen temperature appears to be uniformly large (that is, independently of $\delta B/B_0$), in the sense that most of the potential drop, due to the electric field, is transformed into heat. This means that thanks to their larger Larmor radius, O$^+$ ions are able to gain a large fraction of the potential drop. This is converted into disordered motion by the chaotic dynamics in the quasi-neutral sheet, while the influence of turbulence is evident mostly in the outer part of the simulation box, see \ref{T_O_+}.
 Once again, the comparison of proton and oxygen dynamics shows that the former interact more with turbulence than the latter. To some degree, we can say that the turbulent fluctuations are averaged along the gyroorbit, a behaviour wich is also found in the studies of plasma transport in the presence of turbulence (e.g., \cite{ZimbardoEA06, PommoisEA07}).
 Finally, when realistic ratios of oxygen to proton are assumed, it is found that a bifurcated current sheets is obtained even in absence of magnetic fluctuations. This suggests that the presence of oxygen ions as the main current carriers can help to explain the observations of bifurcated current sheets in the magnetotail. On the other hand, a complete description of the magnetotail current should take into account the electron contribution (e.g.,\cite{GrecoEA07}), as we plan to do in a future work.

\acknowledgements
This work was partially supported by the Italian Space Agency, contract ASI n. I/015/07/0 "Esplorazione del sistema solare".

\bibliographystyle{agu}

\end{article}
\end{document}